\documentclass[aps,twocolumn,superscriptaddress,preprintnumbers]{revtex4}
\usepackage{graphicx}
\usepackage{amsmath,amssymb,bm}
\usepackage[mathscr]{eucal}
\usepackage[colorlinks,linkcolor=blue,citecolor=blue]{hyperref}

\newif\ifarxiv
\arxivfalse

\begin		{document}

\def\Arxiv      #1 [#2]{\href{http://arxiv.org/abs/#1}{{\tt arXiv:#1 [#2]}}\,}

\title
    {
    Jet quenching in strongly coupled plasma
    }
\author{Paul~M.~Chesler}

\affiliation
    {%
Department of Physics, 
Harvard University, 
Cambridge, MA 02138, USA 
    }%

\author{Krishna Rajagopal}

\affiliation
    {%
Center for Theoretical Physics, 
Massachusetts Institute of Technology, 
Cambridge, MA 02139, USA 
    }%

\date{February 26, 2014}

\begin{abstract}
We present calculations in which an energetic light quark shoots through
a finite slab of strongly coupled ${\cal N}=4$ supersymmetric Yang-Mills (SYM) 
plasma, with thickness $L$, focussing on what comes out on the other side.  We find that even when the ``jets''
that emerge from the plasma have lost a substantial fraction of their energy
they look in almost all respects  like  ``jets'' in vacuum with the same reduced energy.
The one possible exception is that the opening angle of the ``jet''
is larger
after passage through the slab of plasma than before. 
Along the way, we obtain a fully geometric characterization of energy loss in 
the strongly coupled plasma and
show that $dE_{\rm out}/dL \propto L^2/\sqrt{x^2_{\rm stop}-L^2}$, where
$E_{\rm out}$ is the energy of the ``jet'' that emerges from the slab of plasma and $x_{\rm stop}$
is the (previously known) stopping distance for the light quark in an infinite volume of plasma.  
\end{abstract}

\preprint{MIT-CTP-4527}


\maketitle


\section{Introduction and Conclusion}

One of the striking early discoveries made by analyzing heavy ion collisions
at the LHC is that when a hard parton with an initial energy of a few hundred GeV loses a significant
fraction of its energy as it plows through a few fm of the hot (temperature $T$ such that $\pi T$ is of order
1 GeV) strongly coupled plasma produced in the collision, the jet that emerges looks remarkably similar
to an ordinary jet produced in vacuum 
with the same, reduced, energy.  This is so even though the jet that emerges from the collision has manifestly
been substantially modified by its propagation through the plasma --- it has lost a substantial fraction of
its energy.  
The ``lost'' energy is found in many soft particles, with momenta comparable to $\pi T$, produced at large
angles relative to the jet.  It is as if the lost energy has become a little more, or a little
hotter, plasma.  These qualitative observations were first made in Refs.~\cite{Aad:2010bu,Chatrchyan:2011sx,Chatrchyan:2012nia}, in particular in Ref.~\cite{Chatrchyan:2011sx}.  
Subsequent 
measurements have quantified these observations further, 
and in particular have quantified what ``remarkably similar'' means
by measuring various small differences between the quenched jets and vacuum jets with the same
energy as the quenched jets~\cite{Chatrchyan:2012gt,Chatrchyan:2012gw,Aad:2012vca,Aad:2013sla,Chatrchyan:2013kwa}.  Here we shall focus on the original qualitative observation, which
remains striking.  We shall argue that this phenomenon is natural in a strongly
coupled gauge theory by doing a calculation in which we shoot a light quark ``jet'' in ${\cal N}=4$ SYM
theory through a slab of strongly coupled plasma and looking at what comes out on the other side.  
Our conclusion can only be qualitative for the simple reason that there are no jets in ${\cal N}=4$ SYM 
theory~\cite{Hofman:2008ar,Hatta:2008tx,Chesler:2008wd}.
The light quark ``jet'' in this theory should not be compared quantitatively to a jet in QCD.  Nevertheless,
we shall find that even when one of these ``jets'' loses a substantial fraction of its energy
as it propagates through a slab of plasma, it emerges looking precisely like a ``jet'' with the 
same (reduced) energy and same (increased) opening angle
would look in vacuum.

The conclusion that we reach is consistent with conclusions (also qualitative) reached
by analyzing the quenching of a beam of gluons by strongly coupled ${\cal N}=4$ SYM 
plasma~\cite{Chesler:2011nc}.
In a different sense, weak-coupling analyses of the quenching of a high energy parton by 
a slab of weakly coupled plasma at some constant $T$  (see Ref.~\cite{Armesto:2011ht} and 
references therein)
are also antecedents of our calculation, although the physics there is quite different since energy is lost,
at least initially, to radiated gluons with momenta $\gg \pi T$ that are nearly collinear with the initial parton.
More recent weak-coupling analyses, beginning with Ref.~\cite{CasalderreySolana:2010eh}, have shown how
the energy lost from jets can go to large angles;
for a recent review of weak-coupling analyses of jet quenching, see Ref.~\cite{Mehtar-Tani:2013pia}.

\begin{figure}[t]
\includegraphics[scale = 0.3]{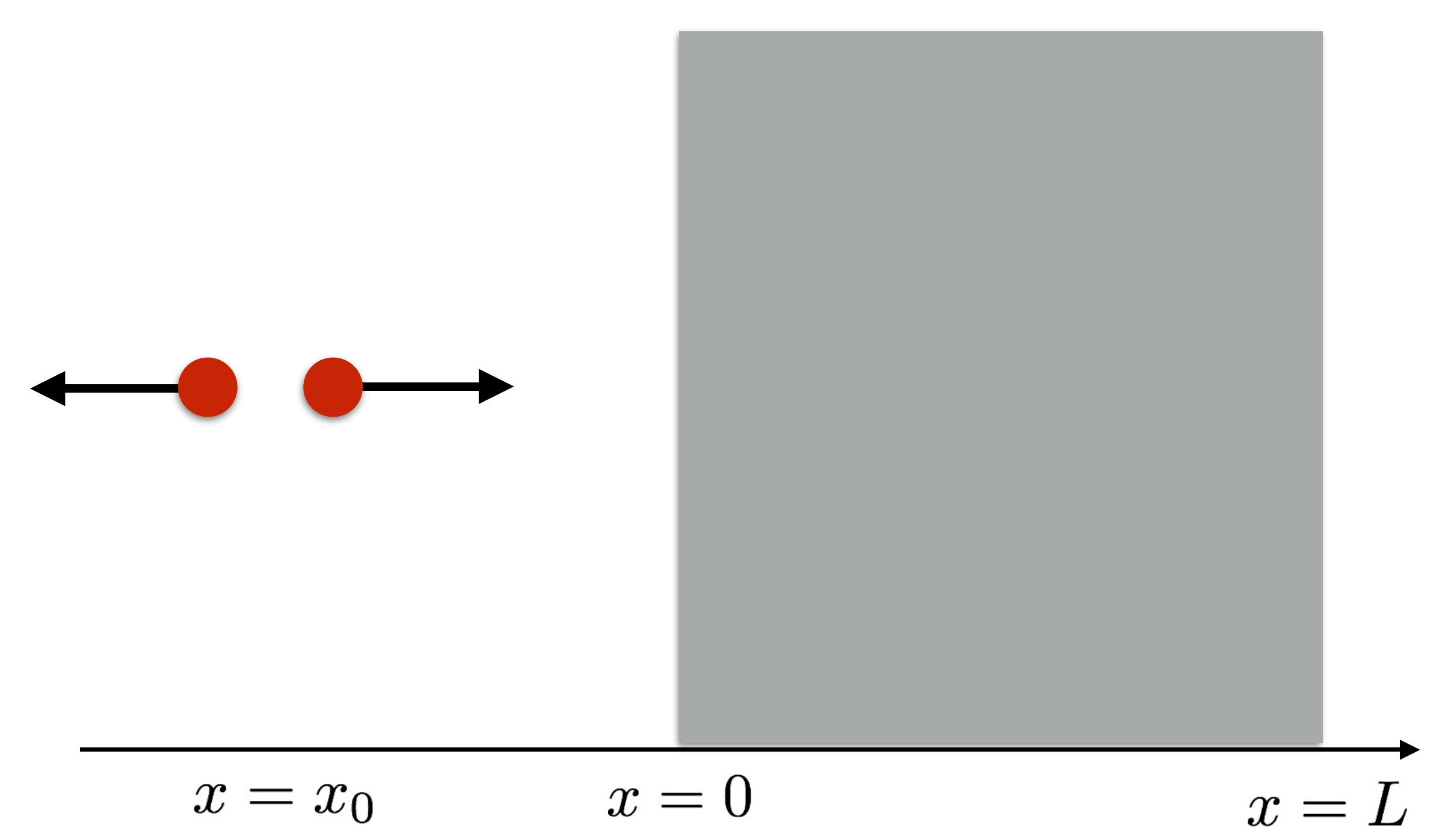}
\caption{A cartoon of the setup of our problem.  A pair of quarks (red circles) are 
created at time $t = 0$ at $ x = x_0$ with momentum in the $\pm x$ direction.  The shaded region shows 
the slab of plasma.  The right-moving quark impacts the plasma at time $t  \approx |x_0|$ and 
exits the plasma at time $t \approx |x_0| + L$.
\label{fig:slab}
} 
\end{figure}

\section{Setup and string dynamics}
\label{sec:stringdynamics}

The setup of our problem is as follows.  We consider an energetic 
pair of massless quarks created in vacuum at $x = x_0 < 0$ at time $t = 0$.
The quarks subsequently move apart in the $\pm x$ direction.  The slab of strongly coupled
${\cal N}=4$ SYM plasma occupies the region $x \in (0,L)$ and $y,z \in (-\infty,\infty)$.
Hence the right-moving quark impacts the plasma at time $t  \approx |x_0|$ and exits the 
plasma at time $t \approx |x_0| + L$.  This 
setup is depicted in the cartoon in Fig.~\ref{fig:slab}.  Although the most relevant case is
that with $x_0=0$, since in a heavy ion collision the energetic quark is produced within
the plasma rather than being incident upon it from outside, it will nevertheless be advantageous to analyze
the case with $x_0<0$ first. One reason for this is that
it allows us to compute the expectation value of the stress tensor $\langle T^{\mu\nu}\rangle$
of the incident ``jet'', between $x=x_0$ and $x=0$.  Our
particular interest is then the computation of
$\langle T^{\mu \nu} \rangle$ in the ``out" region $x \to \infty$. 
How does the presence of the slab alter $\langle T^{\mu \nu}\rangle$?  How is the shape of the jet altered by the slab?
How does the slab change the total energy and momentum of the jet?  

In this Section, we shall discuss the dual gravitational description of the above process
in terms of the dynamics of energetic strings in an asymptotically AdS$_5$ geometry.
We begin by constructing the string solutions, and then use the gravitational
description that they provide to obtain fully geometric characterizations of
the energy loss experienced by the light quark traversing the slab of plasma,
the stopping distance for the light quark if the slab of plasma were so thick
that the energetic quark does not make it through, and the change in the opening
angle of the  ``jet'', namely the boosted beam of energy around the light quark.  
We derive analytic expressions for the rate of energy loss in the (unphysical)
case of a light quark that has propagated a long distance between its creation
and the moment when it enters the slab of plasma and for the (more realistic,
in the context of heavy ion collisions) case of a light quark that enters the
slab of plasma immediately after it is produced.  In both cases, we find
a Bragg peak, which is to say that we find that the rate of energy loss is
greatest for those light quarks that are fully stopped by the plasma, and is 
greatest as the distance they have travelled approaches their stopping distance.
In Section~\ref{sec:BoundaryInterpretation} we compute the angular distribution
of power radiated by the ``jet'' that escapes the slab of plasma, confirming that
our gravitational calculation of the energy loss in terms of the energy of the segment
of string that emerges from the plasma is indeed the calculation of the energy lost
by the ``jet''.  And, we confirm that the shape of the ``jets'' that emerge from the plasma
is the same as that of the ``incident'' jets, even when they have lost a substantial
fraction of their energy and even when their opening angle has increased substantially.
 
%
%

According to gauge/string duality, a quark-gluon plasma is dual to a black hole geometry \cite{Witten:1998qj}.  We model the 
above-horizon geometry corresponding to the slab of plasma with a constant
temperature $T$ with the metric
\begin{equation}
\label{eq:metric}
  {  ds^2 = \frac{1}{u^2}
    \left [-f(x,u) \, dt^2 + d \bm x^2 + \frac{du^2}{f(x,u)} \right ] ,}
\end{equation}
where $f(x,u) = h(u)$ for $0 < x < L$ with $h(u) \equiv 1 - u^4/u_h^4$ and 
$f(x,u) = 1$ otherwise.  The temperature of the slab of plasma is related to the horizon
radius $u_h$ via $u_h = 1/\pi T$.  The boundary of the geometry is located at radial coordinate $u = 0$.
While this model of the black hole geometry is unrealistic near the vacuum/plasma interfaces at $x = 0$ and
$x = L$, in exactly the same sense that it is unphysical to have a slab of plasma at constant nonzero
temperature sitting calmly with vacuum next to it rather than exploding,
in the $T L \gg 1$ limit interface effects are negligible on the dynamics of the propagating quark 
compared to bulk effects accumulated propagating through the plasma.  

The addition of a massless quark to the boundary QFT state is equivalent to adding a falling string 
to the geometry \cite{Karch:2002sh}.  The dynamics of the string are governed by the 
Nambu-Goto action
\begin{equation}
\label{eq:nambugoto}
S = -T_0 \int d \tau d \sigma \sqrt{-g}
\end{equation}
where the string tension $T_0 \equiv \frac{\sqrt{\lambda}}{2 \pi}$ with $\lambda$ the 't Hooft coupling, $\tau$ and $\sigma$
are worldsheet coordinates,
$g \equiv \det g_{ab}$, $g_{ab} \equiv \partial_a X \cdot \partial_b X$ is the string
worldsheet metric and $X^M = \{t(\tau,\sigma),x(\tau,\sigma),0,0,u(\tau,\sigma)\}$ are the string embedding functions.

Upon suitably fixing worldsheet coordinates, the string equations of motion 
can be expressed in terms of the canonical worldsheet densities $\pi^\tau_M$ 
and fluxes $\pi^\sigma_{M}$ 
\begin{subequations}
\label{eq:wscurrents}
\begin{align}
\label{eq:wsenergy}
\pi^0_M &= - T_0 { \textstyle \frac{G_{MN}}{\sqrt{-g}}} \left [ (\dot X \cdot X') X'^N - (X')^2 \dot X^N \right ],
\\
\pi^\sigma_M &= - T_0 {\textstyle \frac{G_{MN}}{\sqrt{-g}}} \left [ ( \dot X \cdot  X') \dot X^N - (\dot X)^2  X'^N \right ],
\end{align}
\end{subequations}
where $G_{MN}$ is the metric (\ref{eq:metric}) and 
$\dot {} \equiv \partial_\tau$ and $' \equiv \partial_\sigma$.
In terms of these quantities the equations of motion read
\begin{equation}
\partial_\tau \pi^0_0 + \partial_\sigma \pi^\sigma_0 = 0,
\label{eq:stringeqm}
\end{equation}
which encodes
worldsheet energy conservation.

Following Refs.~\cite{Chesler:2008uy, Chesler:2008wd} we 
model the creation of a pair of massless quarks at $x = x_0$ by a string created at the point
\begin{equation}
\label{eq:create}
X^M_{\rm create} = \{0,x_0,0,0,u_0\}.
\end{equation}
The string subsequently expands into a finite size object as time progresses
with endpoints moving apart in the $\pm x$ directions.
Open string boundary conditions require the string endpoints to move at the speed of light in the bulk with 
the endpoint velocity transverse to the string.  (We use standard open string boundary conditions
throughout; other boundary conditions have also been considered~\cite{Ficnar:2013wba,Ficnar:2013qxa}.)
Since the $x$ position of the endpoints
corresponds approximately to the position of the quarks in the QFT~\cite{Chesler:2008uy}, we consider strings 
whose endpoint velocities in the $\pm x$ directions are asymptotically close to the speed of light
and which therefore fall only slowly in the radial direction.  
We will confirm below that such strings have asymptotically high energy $E_{\rm string} \to \infty$
and have small $\sqrt{E^2_{\rm string}-p^2_{\rm string}}/E_{\rm string}$,
meaning that they correspond in the dual QFT to excitations that propagate at nearly the
speed of light and that have a small opening angle, like QCD jets with a small angular extent
in momentum space
$\sim m_{\rm jet}/E_{\rm jet}$, with the jet mass $m_{\rm jet}\equiv \sqrt{E^2_{\rm jet}-p^2_{\rm jet}}$.

As the strings have finite tension, the $E_{\rm string} \to \infty$ limit 
is generically realized by strings that expand at nearly the speed of light,
meaning that the string profile must be approximately that of 
an expanding filament of null dust.  Indeed, null strings satisfy $g(X_{\rm null}) = 0$
and from (\ref{eq:wsenergy}) have divergent energy density.  As we detail below, solving the string equations perturbatively 
about a null configuration is tantamount to solving them 
using geometric optics,
with perturbations propagating on the string worldsheet along 
null geodesics.

Since null strings satisfy $g(X_{\rm null}) = 0$ they minimize the Nambu-Goto action (\ref{eq:nambugoto})
and are exact, albeit infinite energy, solutions to the string equations of motion (\ref{eq:stringeqm}) %
\footnote
  {
  This statement requires further clarification since the string equations are singular
  when $g = 0$.  To make this statement more precise, 
  substitute the expansion (\ref{eq:expansion}) into the string equations of motion
  (\ref{eq:stringeqm}).  Then $g(X) = O(\epsilon)$ is finite.  It is then easy to show that in the 
  $\epsilon \to 0$ limit  
  the resulting non-singular equations of motion for $X_{\rm null}$ are satisfied if $g(X_{\rm null}) = 0$.
  }.
To obtain finite energy solutions we expand the string embedding functions about a null string solution
\begin{equation}
\label{eq:expansion}
X^M = X_{\rm null}^M + \epsilon \, \delta X_{(1)}^M + \epsilon^2 \delta X_{(2)}^M + \dots,
\end{equation}
where $X_{\rm null}^M$ is a null string expanding everywhere at the speed
of light and
where $\epsilon$ is a bookkeeping parameter (which we shall see below 
via (\ref{eq:stringenergy})
is related to the string energy via $E_{\rm string} \sim 1/\sqrt{\epsilon}$)
that we shall initially treat as small for the purposes of 
organizing the non-linear corrections to the null string solution
but
that must in the end be set to $\epsilon=1$.
We choose worldsheet coordinate $\tau = t$ and define $\sigma$ by the conditions
$\partial_t X_{\rm null} \cdot \partial_\sigma X_{\rm null} = 0$ and $\delta X_{(n)}^M = \{0,\delta x_{(n)},0,0,0\}$.
We then solve the string equations (\ref{eq:stringeqm}) perturbatively in powers of $\epsilon$.  
The first step is constructing null string solutions.

\subsection{Constructing null strings}

Null strings can be constructed out of a congruence of null geodesics.
Each geodesic in the congruence can be labeled by $\sigma$
and parameterized by time $t$.
The null string can then be written
\begin{equation}
\label{eq:nullstring}
X_{\rm null}^M = \{t,x_{\rm geo}(t,\sigma),0,0,u_{\rm geo}(t,\sigma)\},
\end{equation}
where $x_{\rm geo}$ and $u_{\rm geo}$ satisfy the null geodesic equations
\begin{subequations}
\label{eq:2ndordergeo}
\begin{align}
\frac{\partial}{\partial t} \left (\frac{1}{f} \frac{\partial x_{\rm geo}}{\partial t} \right ) + \frac{1}{2 f}\left ( 1 + \frac{1}{f^2} \left (\frac{\partial u_{\rm geo}}{ \partial t} \right )^2 \right ) \frac{\partial f}{\partial x} 
&= 0,
\\
-f + \left (\frac{\partial x_{\rm geo}}{\partial t} \right )^2 + \frac{1}{f}  \left (\frac{\partial u_{\rm geo}}{ \partial t} \right )^2 
&=0,
\end{align}
\end{subequations}
and
the constraint $\partial_t X_{\rm null} \cdot \partial_\sigma X_{\rm null} = 0$.  
The required initial condition for the congruence is $X^M_{\rm null}|_{t = 0} = X^{M}_{\rm create}$.

For our piecewise constant (in $x$) geometry  the null geodesic equations (\ref{eq:2ndordergeo}) may be integrated once to yield
\begin{subequations}
\label{eq:geos}
\begin{align}
\label{eq:geosx}
\frac{\partial  x_{\rm geo}}{\partial t} &= \frac{f}{\xi}, \ 
\\ \label{eq:geosu}
\frac{\partial u_{\rm geo}}{\partial t} &=  \frac{f \sqrt{\xi^2 - f}}{\xi},
\end{align}
\end{subequations}
which yield a null trajectory satisfying
\begin{equation}
\frac{\partial u_{\rm geo}}{\partial  x_{\rm geo}}=\sqrt{\xi^2-f},
\end{equation}
where the constant of integration $\xi(\sigma)$ is piecewise time-independent in each interval but discontinuous at each interface:
\begin{equation}
   \xi(\sigma) = \left\{
     \begin{array}{lrl}
       \xi_{\rm in}(\sigma), & x <&0,  \\
       \xi_o(\sigma), & 0 <& x < L, \\
       \xi_{\rm out}(\sigma), & x >& L.
     \end{array}
   \right.
\end{equation}

\begin{figure*}[t]
\includegraphics[scale = 0.59]{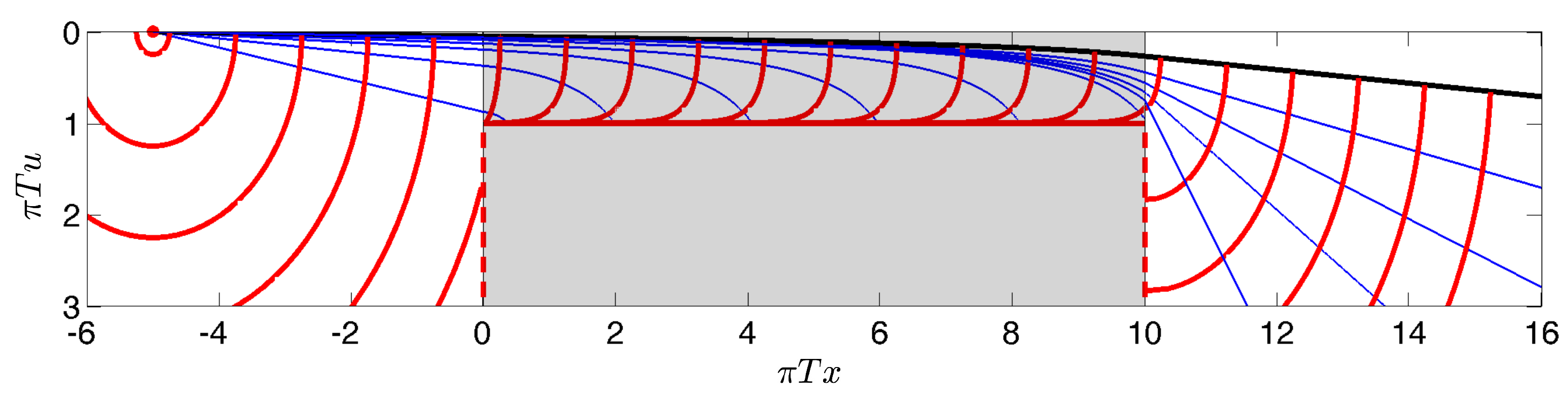}
\caption{A null string.  The string starts off as a point at $x_0 = -5$, $u_0 = 0$
and subsequently expands into a semicircular arc, with its endpoint having $\sigma_*=0.01$.
The small value of $\sigma_*$ ensures that the initial endpoint velocity $dx_{\rm endpoint}/dt=\cos\sigma_*$
is close to the speed of light.
The null string profile is shown (red curves) at times $t=0.25$ through $t=20.25$ in
$\Delta t=1$ increments.  The blue curves are null geodesics propagating along
the string worldsheet at constant values of $\sigma$.  
Energy on the string is transported
along such $\sigma=$constant geodesics,
meaning that the fact that the above-horizon string segment loses energy
as it
propagates through the slab corresponds precisely to the fact that 
within the slab some of
the blue curves fall into the horizon, located at $u= u_h=(\pi T)^{-1}=1$ in our units. 
The string that emerges from the slab
carries only the energy that is transported along those blue curves that emerge.
The string enters the slab at $x=0$ with its endpoint at $u_{\rm in}=0.05$.
The string exits the slab at $\pi T x = 10$ with its endpoint at $u_{\rm out}=0.276$ and having $\tilde\sigma_*=0.0773$.
The energy of the string that exits the slab of plasma is less than that which entered it
by the ratio $E_{\rm out}/E_{\rm in}$, which is 0.643 according to (\ref{eq:ratio}) and 0.57 according
to the $x_0\to-\infty$ approximation (\ref{eq:analytic}).
The string has lost a substantial fraction of its energy while
propagating through the plasma.
After exiting the slab, the string rapidly
approaches a semicircular arc configuration at late times, looking just like a string produced 
in vacuum with the (reduced) energy $E_{\rm out}$ and the (increased) opening angle 
$m_{\rm out}/E_{\rm out} \simeq \sin\tilde\sigma_*$. 
If there really were some way to stabilize interfaces between a slab of plasma and vacuum, 
we expect that
the strings would be connected at the interfaces via vertical segments indicated schematically by
the dashed red lines.  
\label{fig:fallingstring}
} 
\end{figure*}

In the region $x < 0$ where $f = 1$, the geodesic equations (\ref{eq:geosx}) and (\ref{eq:geosu}) may easily be integrated to yield
\begin{equation}
\label{eq:zeroTleft}
x_{\rm geo} = t \cos \sigma + x_0, \ 
u_{\rm geo} = t \sin \sigma + u_0.
\end{equation}
The geodesic is specified by $x_0$, $u_0$ and the
parameter  $\sigma$, which is simply the angle of the geodesic trajectory in the
half-plane $(x,u>0)$.
From the geodesic equations (\ref{eq:geosx}) and (\ref{eq:geosu}) we therefore identify 
\begin{equation}
\label{eq:xileft}
\xi_{\rm in}(\sigma) = \sec \sigma.
\end{equation}
The minimum $\sigma_* \equiv {\rm min}(\sigma)$ corresponds to the endpoint trajectory in the $+x$ direction;
this trajectory is given by $(u_{\rm geo}-u_0)/(x_{\rm geo}-x_0)=\tan \sigma_*$, see
Fig.~\ref{fig:fallingstring}.  We shall see below that to the extent that
the energy of the string is dominated by the energy density near its endpoint,
a string whose endpoint follows this vacuum trajectory has $m\equiv \sqrt{E^2-p^2}=E \sin\sigma_*$,
meaning that $\cos\sigma_*$ is the velocity of the corresponding excitation in the QFT
and $\sin\sigma_*$ is its opening angle.  In Section~\ref{sec:BoundaryInterpretation} we will compute
the angular distribution of the energy of the ``jet'' in the boundary theory that is described by
the string.  We shall see that equating $\sin\sigma_*$ with the opening angle of the ``jet'' is
a good approximation as long as $\sin\sigma_*$ is small.  It is, however, only an approximation
because the relation  $m/E = \sin\sigma_*$ only becomes accurate for the component of the ``jet'' that
is described by the energy density of the string in the vicinity of the endpoint of the string and although the energy
density of the string deep within the bulk contributes less it
does in fact contribute.

The discontinuities in $\xi$ at the $x = 0$ and $x = L$ interfaces can easily be worked out from the second order geodesic equations (\ref{eq:2ndordergeo}).
A geodesic labeled by $\sigma$ passes through $x = 0$ at time and radial coordinate
\begin{align}
\label{eq:tinuin}
& t_{\rm in} = -x_0 \sec \sigma, & u_{\rm in} = -x_0 \tan \sigma + u_0.  
\end{align}
From the second order geodesic equations (\ref{eq:2ndordergeo})
it follows that for $0 < x < L$ the parameter $\xi_o$ is given by
\begin{equation}
\label{eq:xi0}
\xi_o^2 =   \frac{\xi_{\rm in}^2 h(u_{\rm in})}{\xi_{\rm in}^2 + (1- \xi_{\rm in}^2 ) h(u_{\rm in})^2}.
\end{equation} 
The geodesic equation (\ref{eq:geosx}) in the region $0 < x < L$ 
is solved by 
\begin{align}
\label{eq:plasmageo}
x_{\rm geo} = {\textstyle \frac{u_h^2}{u_{\rm in}} } \, _2 F_1\big({\textstyle \frac{1}{4},\frac{1}{2};\frac{5}{4};\frac{u_h^4}{ \zeta \,u_{\rm in}^4} }\big)
{\textstyle -\frac{u_h^2}{u_{\rm geo}}} \, _2 F_1\big({\textstyle \frac{1}{4},\frac{1}{2};\frac{5}{4};\frac{u_h^4}{ \zeta\, u^4_{\rm geo} } }\big),
\end{align}
where 
\begin{equation}
\label{eq:zetadef}
\zeta \equiv \frac{1}{1 - \xi_o^2 },
\end{equation}
and $_2 F_1$ is the Gauss hypergeometric function.  The solution $u_{\rm geo}$ to (\ref{eq:geosu})
can be expressed in terms of elliptic functions and will not be 
written here.

Likewise, a geodesic labeled by $\sigma$ will pass through $x = L$ at some time $t_{\rm out}(\sigma)$ and at some radial coordinate $u_{\rm out}(\sigma)$.
In terms of these quantities the second order geodesic equations (\ref{eq:2ndordergeo}) imply
\begin{equation}
\label{eq:xiout}
\xi_{\rm out}^2 =  \frac{\xi_o^2 h(u_{\rm out})^2}{\xi_o^2 (h(u_{\rm out})^2  -1 ) +  h(u_{\rm out})}.
\end{equation} 
In the region $x > L$ where again $f = 1$ the solutions to the geodesic equations read
\begin{align} 
\label{eq:zeroTright}
\!\!\! x_{\rm geo} = (t {-} t_{\rm out}) \cos \tilde \sigma + L, \
u_{\rm geo} = (t {-} t_{\rm out}) \sin \tilde \sigma + u_{\rm out}.
\end{align}
Via Eq.~(\ref{eq:geosx}) the function $\tilde \sigma(\sigma)$ is given by 
\begin{equation}
\label{eq:xiright}
\xi_{{\rm out}}(\sigma) = \sec \tilde \sigma(\sigma).
\end{equation}

Fig.~\ref{fig:fallingstring} shows 
a few null geodesics (blue curves) which make up a congruence specified by 
$x_0 = -5$, $L = 10$, $u_0 = 0$ and describe the propagation of
a null string (red curves).  Our choice of units here and in what follows is set by $\pi T = 1$.
The trajectory of the endpoint 
moving in the $+x$ direction is given by $\sigma \equiv \sigma_* = 0.01$.  
The small value of $\sigma_*$ ensures that the initial endpoint velocity 
$dx_{\rm endpoint}/dt = 1/\xi_{\rm in}(\sigma_*) = \cos(\sigma_*)$ is close to the speed of light.

Before the string passes through $x = 0$, 
the geodesics (\ref{eq:zeroTleft}) 
and the null embedding functions (\ref{eq:nullstring}) imply that 
the null string profile is given by the expanding 
semi-circular arc 
\begin{equation}
\label{eq:inarc}
-t^2 + (x_{\rm geo} - x_0)^2 + (u_{\rm geo} - u_0)^2 = 0.
\end{equation}
After the string has passed through $x = 0$ into the black hole slab, its profile is given by 
\begin{align}
\nonumber
x_{\rm geo}(t,\sigma) = & \xi_o(\sigma)(t - t_{\rm in}(\sigma))\ + \ x_{\rm trailing} \! \left (\sigma,u_{\rm geo}(t,\sigma) \right ) \\
- \ & x_{\rm trailing} \! \left (\sigma,u_{\rm in}(t,\sigma) \right ),
\end{align}
where $x_{\rm trailing}$ satisfies 
$\partial x_{\rm trailing}/\partial u_{\rm geo} = -\sqrt{\xi_o^2 - h}/h.$
For $\xi_o = 1$, $x_{\rm trailing}$ 
is the null limit of the trailing string profile of Refs.~\cite{Herzog:2006gh,Gubser:2006bz}.
Indeed, geodesics that propagate farthest originate from near the string's endpoint
and have $\xi_o(\sigma) \approx \xi_o(\sigma_*) \approx 1$.
After the string has exited the black hole slab at $x = L$ the  geodesics (\ref{eq:zeroTright}) 
and the null embedding functions (\ref{eq:nullstring}) imply that 
the null string profile is given by 
\begin{equation}
\label{eq:outarc}
-(t-t_{\rm out})^2 + (x_{\rm geo} - L)^2 + (u_{\rm geo} - u_{\rm out})^2 = 0.
\end{equation}

Comparing (\ref{eq:outarc}) and (\ref{eq:inarc}) we see that at asymptotically late times the string profile for $x > L$ is an expanding 
semicircular arc, {\it precisely} as it was for $x<0$.  This is a consequence of the 
fact that as viewed from $x \gg L$ the ``aperture" at $x = L$, $u \in (0,u_h)$ is effectively a point-source 
emitter for null geodesics in the $(x,u)$ plane
 just as the point $x=x_0$, $u=u_0$ was.  Therefore,
other than the fact that endpoint on the right falls with angle 
\begin{equation}
\label{eq:tildesigmastar}
\tilde \sigma_* \equiv \tilde \sigma(\sigma_*)>\sigma_*,
\end{equation}
the null string profile for $x >L$ at late times is the same as that for $x < 0$.  
In other 
words, the net effect of the slab on the null string is simply that the endpoint falls into the bulk
at a faster rate than it did before impacting the slab.

The implication of the result at which we have arrived is that in the QFT
the ``jet'' that emerges from the slab of strongly coupled plasma
has a larger $m/E$ and a larger opening angle than the ``jet'' that entered the slab.
We will determine the increase in the opening angle
of the ``jet'' more precisely in Section~\ref{sec:BoundaryInterpretation},
but as long as the jets remain narrow it is a good approximation to equate
the increase in $\sin\tilde\sigma_*$ relative
to $\sin\sigma_*$ with the increase in $m/E$ and the increase in the opening angle.

With the exception of the increase in $m/E$ and the decrease in $E$ --- see below --- the
``jet'' that emerges looks precisely the same as that which entered. In
particular, it looks precisely the same as a ``jet'' in vacuum prepared
with a larger $m/E$ and a smaller $E$.  This conclusion
comes directly from seeing that the shape of the string
is the same after exiting the plasma as before entering it, and this
in turn is a result that is obtained completely geometrically, as in Fig.~\ref{fig:fallingstring}.
This central conclusion of our study resonates strongly with the observations
of the highest energy jets produced in heavy ion collisions at the LHC with
which we began.

\begin{figure*}[t]
\includegraphics[scale = 0.475]{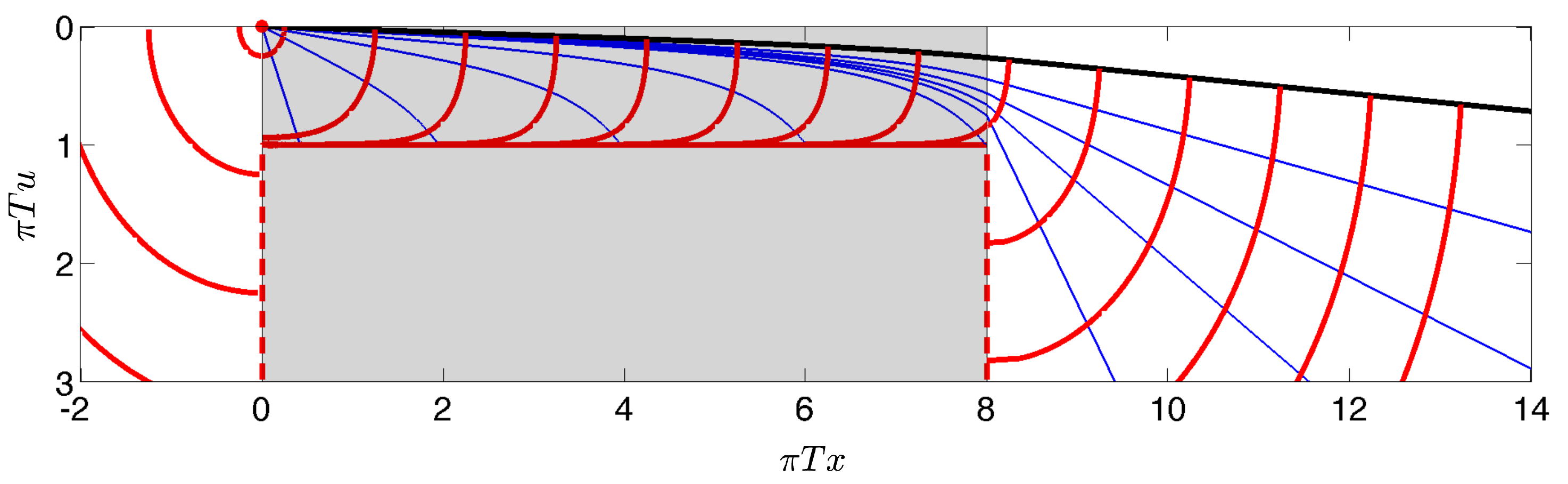}
\caption{As in Fig.~\ref{fig:fallingstring}, except here the slab has thickness $L=8/(\pi T)$ and the
 quark is produced next to the slab 
at $x_0=-10^{-3}$ with $u_0= 0$ and $\sigma_*=0.025$.  It emerges
from the slab at $\pi T x=8$ with $u_{\rm out}=0.267$ and $\tilde\sigma_*=0.0769$.
As in Fig.~\ref{fig:fallingstring}, after exiting the slab of plasma the string
rapidly approaches a semicircular arc configuration.  
Using (\ref{eq:ratio}), $E_{\rm out}/E_{\rm in}=0.757$. 
The approximation (\ref{eq:Eratio}) yields $E_{\rm out}/E_{\rm in}=0.780$. 
So, as in Fig.~\ref{fig:fallingstring} the string has lost a substantial fraction of its energy
in the plasma and yet emerges looking just like a string produced in vacuum with energy $E_{\rm out}$.
We shall see in Section II.C that, under certain assumptions, this string describes
a ``jet'' with  an incident energy $E_{\rm in}=87.0 \sqrt{\lambda} \,\pi T$ and, consequently, an
outgoing ``jet'' that emerges from the slab with energy  $E_{\rm out}=65.9 \sqrt{\lambda} \,\pi T$. 
\label{fig:morerelevant}
} 
\end{figure*}

It is reasonable to ask whether the conclusion that we have just reached depends on the fact
that we created the energetic string well to the left of the slab, allowing the string
to propagate some distance in AdS before entering the slab.  In a heavy ion collision, after all,
the high energy parton rapidly finds itself in the strongly coupled matter produced in the collision.
We show in Fig.~\ref{fig:morerelevant} that we reach the same conclusion upon considering a
case in which the energetic string is produced at $x_0=-10^{-3}$ and immediately enters the slab.
Indeed, the conclusion that the null string that emerges from the slab quickly becomes vacuum-like
in appearance (i.e. quickly becomes semicircular) is completely generic because
it arises directly from the geometric perspective that our holographic calculation provides.

\subsection{First order corrections}

With the null string dynamics worked out, we now turn to the first order 
perturbations $\delta x_{(1)}$ in terms of which the worldsheet energy density
and flux read
\begin{align}
\label{eq:stringenergy}
 \pi^0_0 &=  
 -\frac{ T_0\, \xi\, \partial_\sigma u_{\rm geo}}{u_{\rm geo}^2} \sqrt{   \frac{-\xi}{2\, \epsilon\, f \,\partial_t \delta x_{(1)}} }
 , 
&\pi^\sigma_0 &= 0,
\end{align}
up to order $\sqrt{\epsilon}$ corrections.
According to the string equation of motion (\ref{eq:stringeqm}) at leading order in $\epsilon$ the equation of motion 
for $\delta x_{(1)}$ is simply $\partial_t \pi^0_0 = 0$ so $\pi^0_0$ is time independent and energy 
is simply transported along $\sigma = \rm const.$ geodesics, i.e. along
the blue curves in Figs.~\ref{fig:fallingstring} and \ref{fig:morerelevant}.    
This observation will play a critical role below when
we consider energy loss on the string worldsheet.

Substituting Eqs.~(\ref{eq:zeroTleft}) and (\ref{eq:xileft}) into (\ref{eq:stringenergy}) we find that in the $x < 0$ 
region $\delta x_{(1)}$ must satisfy
\begin{equation}
\partial_t^2 \delta x_{(1)} + 
\frac{2 (t \sin \sigma - u_0)}{t ( t \sin \sigma + u_0)}
\partial_t  \delta x_{(1)} = 0.
\end{equation}
The solution reads 
\begin{equation}
\delta x_{(1)} = \phi(\sigma)  + 
\frac{\sin \sigma  \left(3 t \sin \sigma  \, (t \sin \sigma {+}u_0)+u_0^2\right)}{3 (t \sin \sigma +u_0)^3}
\psi (\sigma ), 
\end{equation}
where $\phi(\sigma)$ and $\psi(\sigma)$ are arbitrary functions.  The condition that the endpoint moves at the speed 
of light requires $\psi(\sigma_*) = 0$.  A simple calculation then yields
\begin{equation}
\label{eq:zeroTenergy}
\pi^0_0 = -T_0 \csc^2 \sigma
\sqrt{ \frac{ \csc 2 \sigma \sin \sigma}{\epsilon \,  \psi(\sigma)}} 
+ O(\sqrt{\epsilon}).
\end{equation}
The solution $\delta x_{(1)}$ in the regions $0 < x < L$ and $x >L$ can then be obtained by 
solving Eq.~(\ref{eq:stringenergy}) for $\partial_t \delta x_{(1)}$ with $\pi_0^0$ given above by (\ref{eq:zeroTenergy}) and integrating in time.

\subsection{Worldsheet energy loss and stopping distance}

We now turn to energy loss in the slab, remembering the geometric intuition from Figs.~\ref{fig:fallingstring} 
and \ref{fig:morerelevant} that energy propagates
along the (blue) null geodesics, with energy loss corresponding to blue geodesics falling
into the horizon.  We begin with the extreme case in which all of the incident energy
is lost, which is to say the case in which the string endpoint falls into the horizon and
no string emerges from the slab of plasma.
Clearly, there exists a maximal distance $x_{\rm stop}$ that
the string endpoint can travel through a $L \to \infty$ slab before the string
endpoint and hence the entire string has fallen into the horizon.  
In the dual field theory the stopping distance $x_{\rm stop}$ corresponds
to the distance a jet can penetrate through the plasma before thermalizing~\cite{Gubser:2008as,Chesler:2008uy}.
From (\ref{eq:plasmageo}) we see that the stopping distance is given by
\begin{align}
\label{eq:xstop}
x_{\rm stop} = -u_h \, & _2 F_1\big({\textstyle \frac{1}{4},\frac{1}{2};\frac{5}{4};\frac{1}{ \zeta(\sigma_*)} }\big)
\\ \nonumber
+ &{\textstyle \frac{u_h^2}{u_{\rm in}(\sigma_*)} } \, _2 F_1\big({\textstyle \frac{1}{4},\frac{1}{2};\frac{5}{4};\frac{u_h^4}{ \zeta(\sigma_*)\, u_{\rm in}(\sigma_*)^4} }\big).
\end{align}
In what follows we shall focus on the limit $x_{\rm stop} \gg u_h$
which generically requires $\sigma_* \ll 1$, so the endpoint
trajectory is nearly constant in $u$ before impacting the slab geometry.  
Restricting our attention to strings created near the boundary, we also set $u_0 \to 0$.  This
is not necessary.  As we discuss in Section~\ref{sec:Outlook}, it will be interesting in future
to systematically explore how our results vary as a function of $u_0$ and $\sigma_*$.

We now return to the case of interest in this paper, namely a slab of plasma whose
thickness $L$ is less than $x_{\rm stop}$ meaning that, as in Figs.~\ref{fig:fallingstring} and \ref{fig:morerelevant},
the endpoint of the string and some of the (blue) null geodesics describing a segment of the 
string near its endpoint emerge from the slab of plasma.
Let us define the function
$\sigma_h(x)$, for $0<x<L$,  by the condition that $x_{\rm geo}(t,\sigma_h) = x$
and $u_{\rm geo}(t,\sigma_h) = u_h$. That is, $\sigma_h(x)$ labels
the null geodesic that falls into the horizon at $x$.
From (\ref{eq:plasmageo}) we see that $\sigma_h(x)$ is the solution to
\begin{align}
\label{eq:sigmah}
x = -u_h \, & _2 F_1\big({\textstyle \frac{1}{4},\frac{1}{2};\frac{5}{4};\frac{1}{ \zeta(\sigma)} }\big)\\ \nonumber
+&{\textstyle \frac{u_h^2}{u_{\rm in}(\sigma)} } \, _2 F_1\big({\textstyle \frac{1}{4},\frac{1}{2};\frac{5}{4};\frac{u_h^4}{ \zeta(\sigma)\, u_{\rm in}(\sigma)^4} }\big),
\end{align}
meaning that $\sigma_h(x_{\rm stop})=\sigma_*$.
The energy of the string segment that exits the slab can then be written
as 
\begin{equation}
E_{\rm out}=-\int_{\sigma_*}^{\sigma_h(L)} d\sigma \, \pi^0_0\,.  
\label{eq:outgoingenergy}
\end{equation}
$E_{\rm out}$ is clearly less
than the energy of the string segment that enters the slab, which we shall take to be
\begin{equation}
E_{\rm in} = -\int_{\sigma_*}^{\sigma_h(0)} d\sigma \, \pi^0_0\,, 
\label{eq:incidentenergy}
\end{equation}
because some null
geodesics and therefore some energy has fallen into the horizon between $x=0$ and $x=L$.

To go further, 
we henceforth assume $u_0\to 0$ and $\sigma_*\ll 1$.
In the $\sigma_* \ll 1$ limit we see from (\ref{eq:zeroTenergy}) 
that $\pi_0^0(\sigma)$ becomes highly concentrated in a region $\delta \sigma \sim \sigma_*$ near $\sigma = \sigma_*$.
Expanding 
\begin{equation}
\psi(\sigma) = \psi'(\sigma_*) (\sigma - \sigma_*)  + {\cal O} \left((\sigma - \sigma_*)^2 \right ),
\end{equation}
we obtain from (\ref{eq:zeroTenergy}) the leading order expression for $\pi_0^0$,
\begin{equation}
\label{eq:endptenergy}
\pi^0_0 = \frac{-T_0}{\sigma^2  \sqrt{2 \epsilon \, \psi'(\sigma_*) (\sigma - \sigma_*})}.
\end{equation}
This expression, together with Eqs.~(\ref{eq:sigmah}), (\ref{eq:outgoingenergy}) and (\ref{eq:incidentenergy}),
allows us to compute
$E_{\rm out}/E_{\rm in}$, which is to say
the fractional energy lost by the high energy parton as it traverses
the slab of plasma.
We obtain
\begin{equation}
\label{eq:ratio}
\frac{E_{\rm out}}{E_{\rm in}} = 
\frac{ \hat \sigma_h(0) \left ( \sqrt{ \hat \sigma_h(L)- 1} +  \hat \sigma_h(L)\cos^{-1} \sqrt{{\textstyle \frac{1}{\hat \sigma_h(L)}}} \right )}
{\hat \sigma_h(L) \left ( \sqrt{ \hat \sigma_h(0)- 1} +  \hat \sigma_h(0) \cos^{-1}\sqrt{ {\textstyle \frac{1}{\hat \sigma_h(0)}}} \right )} 
,
\end{equation}
where $\hat \sigma_h(x) \equiv \sigma_h(x)/\sigma_*$.
Although it does not look particularly simple, this expression
is fully explicit. For example, as noted in the captions of both Figs.~\ref{fig:fallingstring} and
\ref{fig:morerelevant}, we can use it to compute $E_{\rm out}/E_{\rm in}$
for the ``jets'' in both these figures.

We shall next describe two contexts in which the expressions (\ref{eq:xstop}) and (\ref{eq:ratio}) simplify considerably.

\subsubsection{A parton incident from $x_0=-\infty$}
\label{sec:largex0}

The first simplifying limit that we shall consider is 
the limit in which we take
$x_0 \to -\infty$ while fixing $u_{\rm in}$ small compared to $u_h$.
As is evident from (\ref{eq:xstoplargex0}) below, this is equivalent to 
keeping $x_{\rm stop}$ finite (but large compared to $u_h$) as $x_0 \to -\infty$.
This limit, which is
not realistic from the point of view of heavy ion collisions, corresponds to considering
an incident parton that has propagated for a long distance before it reaches the
slab of plasma, but that was prepared with such a small initial opening angle that
when it reaches the slab of plasma the size of the cloud of energy density that
it describes is still small.  
In this limit, $\sigma_*=\arctan(u_{\rm in}/|x_0|)$ vanishes
as $|x_0|\to\infty$ at fixed, small, $u_{\rm in}$.   
In the $x_0\to -\infty$  limit,
$\xi_{\rm in} \to 1$ and $\xi_o^2 \to h(u_{\rm in})$ and
the stopping distance (\ref{eq:xstop}) takes the form
\begin{equation}
\label{eq:xstoplargex0}
x_{\rm stop} =  \frac{\sqrt{\pi } \,\Gamma (\frac{5}{4})}{\Gamma (\frac{3}{4})} \frac{u_h^2}{u_{\rm in}} - u_h
+ \frac{ u_h^4}{2 u_{\rm in}^2 x_0} +{\cal O}(x_0^{-2})\ .
\end{equation}
Neglecting transients
at small $x$, Eq.~(\ref{eq:sigmah}) then yields
\begin{equation}
\label{eq:sigmah2}
\hat \sigma_h(L) = \frac{x_{\rm stop} + u_h}{L + u_h}.
\end{equation}
This means that $\sigma_h(L)$ is  ${\cal O}(\sigma_*)$ when $L = {\cal O}(x_{\rm stop})$, from which it follows that 
$\pi_0^0$ in (\ref{eq:outgoingenergy}) may consistently be taken to be given by the near-endpoint 
expression (\ref{eq:endptenergy}).
Substituting (\ref{eq:sigmah2}) into (\ref{eq:ratio}) and taking $L, \ x_{\rm stop} \gg u_h$ we secure the result
\begin{equation}
\label{eq:analytic}
 \frac{E_{\rm out}}{E_{\rm in}} = \frac{2}{\pi } \left [ \sqrt{  \frac{L (x_{\rm stop} - L)}{x_{\rm stop}^2}}  + \cos^{-1} \sqrt{  \frac{L}{x_{\rm stop}}} \right ].
\end{equation}
Taking the derivative of  (\ref{eq:analytic}), we find the energy loss rate
\begin{equation}
\label{EnergyLossEquation}
 \frac{1}{E_{\rm in}} \frac{dE_{\rm out}}{dL} = -\frac{2}{\pi x_{\rm stop}} \sqrt{ \frac{L}{x_{\rm stop} - L}}\,.
\end{equation}
Eqs.~(\ref{eq:analytic}) and (\ref{EnergyLossEquation}) are
our final results for the energy loss in the (unphysical) case in which
$x_0\to -\infty$.
Eq.~(\ref{eq:analytic}) provides a reasonable approximation in the case illustrated in Fig.~\ref{fig:fallingstring},
but it cannot be applied in the case illustrated in Fig.~\ref{fig:morerelevant}.

\subsubsection{A parton produced at fixed $x_0$ whose $x_{\rm stop} \to \infty$}

Since a hard parton produced 
in a heavy ion collision is produced within the same volume in which the
strongly coupled plasma is produced, the calculation in Fig.~\ref{fig:morerelevant}
in which the parton was produced just next to the slab of plasma
is a better caricature than that in Fig.~\ref{fig:fallingstring}.
We therefore do not wish to take the $x_0\to -\infty$ limit. 
Henceforth, we take the $\sigma_*\to 0$ limit
at fixed $x_0$.  We shall see below that $x_{\rm stop} \to \infty$ in this limit.
We continue to assume
that $u_0=0$, which now means that $u_{\rm in}\to 0$ as $\sigma_*\to 0$.
The results we shall derive here in this limit are a good approximation
for small enough $\sigma_*$ 
at any fixed value of $x_0$, in particular for the case in which 
the parton is produced just next to the slab of plasma,
with $x_0$ just to the left of $x=0$ as in Fig.~\ref{fig:morerelevant}.  


With $u_0=0$ and $x_0$ fixed in value, 
we find that $x_{\rm stop}$ in
(\ref{eq:xstop}) takes the form
\begin{equation}
\label{eq:xstop2}
x_{\rm stop} = \frac{u_h\, \Gamma(\frac{1}{4})^2 }{ 4\sqrt{ \pi \sigma_*} } +  (x_0-u_h) + {\cal O}(\sqrt{\sigma_*})
\end{equation}
in the small-$\sigma_*$ limit.
We see 
that if $\sigma_*$ is small enough that
we can neglect the $(x_0-u_h)$ term we have $x_{\rm stop}\gg |x_0 - u_h| = |x_0|+u_h$ and
$\sigma_* = {\cal O}\big (\frac{u_h^2}{x^2_{\rm stop}} \big ) = {\cal O}\big (\frac{1}{(\pi T x_{\rm stop})^2} \big )$.

In the limit in which we take $\sigma_*\to 0$  with $u_0=0$ and $x_0$ fixed 
we can also derive a relationship between $x_{\rm stop}$ and $E_{\rm in}$,
defined  in Eq.~(\ref{eq:incidentenergy}), valid
to leading order in $\sigma_*$.
Note  that the expression (\ref{eq:endptenergy}) tells us that
the energy density on the string is greatest near the string endpoint.
This observation allows us to see that, 
to leading order in $\sigma_*$, 
Eq.~(\ref{eq:incidentenergy}) yields 
\begin{equation}
\label{eq:incidentenergy2}
E_{\rm in} = \frac{\pi T_0}{2 \,\sigma_*^{3/2} \sqrt{2\, \epsilon \,\psi'(\sigma_*)}}.
\end{equation}
Comparing (\ref{eq:incidentenergy2}) and (\ref{eq:xstop2}) and using $T_0 = \frac{\sqrt{\lambda}}{2 \pi}$, $u_h = 1/\pi T$, we find
\begin{equation}
\label{MaxStopping}
x_{\rm stop} = \frac{\pi^{4/3} \mathcal C}{\pi T} \left ( \frac{E_{\rm in}}{\sqrt{\lambda} \pi T } \right )^{1/3},
\end{equation} 
where the dimensionless constant $\mathcal C$ is given by
\begin{equation}
\pi^{4/3} \mathcal C =  \left (\frac{ \epsilon \, 2^5  T^2 \psi'(\sigma_*)}{\pi} \right )^{1/6} {\textstyle \Gamma(\frac{1}{4}) \Gamma(\frac{5}{4}) }.
\end{equation}
The $x_{\rm  stop} \sim E_{\rm in}^{1/3}$ scaling was first obtained in Refs.~\cite{Gubser:2008as,Chesler:2008uy}.   
Numerical simulations of the string equations in Ref.~\cite{Chesler:2008uy} yielded an
estimate for the maximum possible value of $\mathcal C$, for jets whose initial state
is prepared in such a way as to yield the maximal stopping distance for a given $E_{\rm in}$,
namely
${\rm max}(\mathcal C) \approx 0.526$. %
The value $\mathcal C \approx 0.526$ was recently verified analytically in Ref.~\cite{Ficnar:2013wba}.

We have a calculation of $x_{\rm stop}$ in hand in (\ref{eq:xstop2})
and can now
ask  about the value of $E_{\rm in}$. In this context,
we can reread the maximal value of ${\cal C}$ in (\ref{MaxStopping}) as 
telling us the minimum possible $E_{\rm in}$ that
can correspond to a given $x_{\rm stop}$, assuming optimal preparation of the initial state.
Note that if the initial state is prepared well to the left of the slab of plasma as in Fig.~\ref{fig:fallingstring}
then, even if the initial state is prepared optimally at $x=x_0$,
after the string has propagated in vacuum from $x=x_0$ to $x=0$
 its state is not optimally prepared when it enters the plasma, and ${\cal C}$ must be less than $0.526$ in (\ref{MaxStopping}).  
 We can see this by noting that if we start from a case like that in Fig.~\ref{fig:fallingstring} 
and move the point of origin $x_0$ to $x_0=0$, making no other
change and in particular keeping $u_0$ fixed, this does not change $E_{\rm in}$ but it 
decreases $u_{\rm in}$ (to  $u_{\rm in}=u_0$) and increases $x_{\rm stop}$, for example from
12.71 to 17.54 in the case of Fig.~\ref{fig:fallingstring}.  
We see from (\ref{eq:xstop2}) that this $x_0$-dependence of $x_{\rm stop}$ is subleading
in the small-$\sigma_*$ limit: at small enough $\sigma_*$, moving $x_0$ from -5 as in 
Fig.~\ref{fig:fallingstring} to 0 would have a negligible effect on $x_{\rm stop}$.
Nevertheless, the consequence of this formally subleading effect is that
the minimum value of 
$E_{\rm in}/(\pi T)$  in Fig.~\ref{fig:fallingstring} must be greater than that given by (\ref{MaxStopping})
with $x_{\rm stop}=12.71$ and ${\cal C}=0.526$.  

The expression (\ref{MaxStopping}) with ${\cal C}=0.526$ can be applied without caveats
in Fig.~\ref{fig:morerelevant}.  There, $x_{\rm stop}=10.73$ and the minimum possible
incident energy of the ``jet'' in Fig.~\ref{fig:morerelevant}, assuming optimal preparation
of the initial $\psi(\sigma)$, can be read from (\ref{MaxStopping}) with ${\cal C}=0.526$ 
and is given
by $E_{\rm in}/(\pi T)=87.0\,\sqrt{\lambda}$.  If we think of a slab of plasma in which $\pi T\sim 1$~GeV,
the slab in Fig.~\ref{fig:morerelevant} is 1.6~fm thick and the ``jet'' depicted in the Figure, which loses
24.3\% of its energy as it traverses the plasma, has an incident energy of $87.0\,\sqrt{\lambda}$~GeV,
corresponding to a few hundred GeV.

So, we now know that as we take the $\sigma_*\to 0$ limit at fixed $x_0$,
for example for the case in which the parton is produced next to the 
slab of plasma, $x_{\rm stop}$ takes the form (\ref{eq:xstop2}) 
and is related to $E_{\rm in}$
via (\ref{MaxStopping}).
We also continue to assume that 
$x_{\rm stop} \gg u_h$.  Upon making these assumptions, if we consider a slab
of plasma with $L<x_{\rm stop}$ and
$L / x_{\rm stop} = {\cal O}(1)$, Eq.~(\ref{eq:xstop2}) implies
\begin{equation}
\label{eq:sigmah3}
\hat \sigma_h(L) = \left ( \frac{x_{\rm stop}}{L} \right )^2.
\end{equation}
As above in Sec.~\ref{sec:largex0}, for $L = {\cal O}(x_{\rm stop})$ we see $\sigma_h(L) = {\cal O}(\sigma_*)$, from which it again follows that 
$\pi_0^0$ in (\ref{eq:outgoingenergy}) may consistently be taken to be given by the near-endpoint 
expression (\ref{eq:endptenergy}).
Differentiating (\ref{eq:outgoingenergy}) and dividing by $E_{\rm in}$ we then 
obtain  the rate of energy loss
\begin{equation}
\label{eq:energylossrate}
\frac{1}{E_{\rm in}}  \frac{d E_{\rm out}}{dL} = - \frac{4 L^2}{\pi x_{\rm stop}^2 \sqrt{x_{\rm stop}^2 - L^2}}\,.
\end{equation}
Upon integrating (\ref{eq:energylossrate}) we find that in this case the fractional energy loss is given by
\begin{equation}
\label{eq:Eratio}
\frac{E_{\rm out}}{E_{\rm in}} = \frac{2}{\pi} \left [ \frac{L}{x_{\rm stop}} \sqrt{1 - \frac{L^2}{x_{\rm stop}^2}} + \cos^{-1} \frac{L}{x_{\rm stop}} \right ].
\end{equation}
This expression provides a good approximation to
the energy loss in the case illustrated in Fig.~\ref{fig:morerelevant}.
%
We advocate the use of the expressions (\ref{eq:energylossrate}) and (\ref{eq:Eratio})
for
the rate of energy loss in phenomenological modelling of jet quenching
in heavy ion collisions, with $x_{\rm stop}$ in (\ref{eq:energylossrate}) related to 
the initial energy of the energetic parton  and
to the temperature of the plasma at the location of the energetic parton via Eq.~(\ref{MaxStopping}).

\subsubsection{Bragg peak}

A remarkable feature of either (\ref{eq:ratio}) or (\ref{EnergyLossEquation})
or (\ref{eq:energylossrate}) 
is that little energy is lost
until $L \sim x_{\rm stop}$ and then
$dE_{\rm out}/dL$ diverges as $L \to x_{\rm stop}$.  This behavior, which was
first pointed out in Ref.~\cite{Chesler:2008uy}, is in some respects
reminiscent of a Bragg peak.
The geometric origin of the Bragg peak is easy to understand.  For 
$\sigma_* \to 0$ the string energy density (\ref{eq:endptenergy}) is highly concentrated near the string
endpoint and in fact diverges when $\sigma = \sigma_*$, which reflects the fact that open string
boundary conditions require the string endpoint to move at the speed of light.
Assuming that $L > x_{\rm stop}$,
the energy loss rate $dE_{\rm out}/dL = -\pi_0^0(\sigma_h) \,d \sigma_h/dL$ must
therefore grow unboundedly large as the endpoint falls vertically 
into the horizon when it reaches $x=x_{\rm stop}$.  

The boundary theory interpretation of this phenomenon is that the ``jet'' of energy described by the falling
string 
expands in size as it propagates, expanding linearly with distance as it propagates
in vacuum with some constant opening angle and then faster than linearly as it propagates
through the plasma until, when $x\sim x_{\rm stop}$, its size becomes comparable to $1/(\pi T)$ at which 
point it rapidly thermalizes.  It is important to notice that the rapid thermalization sets
in when the size of the ``jet'' becomes comparable to $1/(\pi T)$ which, depending on the way in
which the ``jet'' is prepared, can happen when the velocity of the ``jet'' is still relativistic.  In this
respect the phenomenon is different than the canonical Bragg peak that arises when an electron
losing energy as it passes through matter decelerates to a non-relativistic speed.

\subsubsection{Momentum loss in the slab of plasma}

For completeness, before turning to the boundary interpretation of the ``jets''
whose energy loss we have computed we set up the calculation of how much
momentum they lose as they traverse the slab of plasma.
As was the case with the string energy,
the momentum $P_{\rm out}= \int_{\sigma_*}^{\sigma_h(L)} d\sigma \, \pi^0_x$ of the string segment 
that exits the slab is less than the momentum $P_{\rm in} = \int_{\sigma_*}^{\sigma_h(0)} d\sigma \, \pi^0_x$
of the string segment that 
that entered the slab.
At leading order in $\epsilon$, the momentum density on the string is given by
\begin{equation}
\label{MomentumOverEnergy}
\pi^0_x = -\pi^0_0/\xi.
\end{equation}
To the extent that the energy and momentum of the string are dominated by the contribution
from near the endpoint, (\ref{MomentumOverEnergy}) implies 
that $P_{\rm in}/E_{\rm in}=1/\xi_{\rm in}=\cos\sigma_*$ and $P_{\rm out}/E_{\rm out}=\cos\tilde\sigma_*$,
meaning that $m_{\rm in}/E_{\rm in}=\sin\sigma_*$ and $m_{\rm out}/E_{\rm out}=\sin\tilde\sigma_*$.
This means that we can immediately see from a figure like Fig.~\ref{fig:fallingstring} or \ref{fig:morerelevant}
that $m_{\rm out}/E_{\rm out} > m_{\rm in}/E_{\rm in}$, meaning that the
opening angle of the  ``jet'' that emerges from the slab of plasma is wider than that of the incident  ``jet''. 
The bulk interpretation is that because the string loses energy
as it propagates through the plasma its endpoint is falling more steeply after it emerges than
it was before it entered the plasma.  In both Figs.~\ref{fig:fallingstring} and \ref{fig:morerelevant}
and in all the other examples that we have investigated, the increase in $m/E$ is
greater than the decrease in $E$ meaning that energy loss is accompanied by an increase in $m$.




\section{Boundary Interpretation}
\label{sec:BoundaryInterpretation}

We have computed the amount of energy that the ``jet'' that exits the slab
of strongly coupled plasma has lost as it traverses the slab.  
And, we have seen in the dual gravitational
description that the string that exits the slab of plasma has the same (semi-circular)
shape as the string that was incident on the slab, but that its endpoint emerges
with a value of $\tilde\sigma_*$ that is greater than the $\sigma$ with which it
entered the slab.  In this Section we shall confirm that these observations
imply that the ``jet'' that exits the slab of plasma in the dual field theory
has a larger opening angle than the incident ``jet'' but that other than this
has the same shape. 
%
%
To address these questions we must consider the angular distribution of 
power radiated by the ``jet" that escapes the slab of plasma,
\begin{equation}
\label{eq:powerdef}
\frac{dP_{\rm out}}{d \Omega} \equiv \lim_{|\bm x| \to \infty} |\bm x|^2  \hat  x_i \int dt \, \langle T^{0 i} \rangle, 
\end{equation}
where $\langle T^{\mu \nu} \rangle$ is the expectation value of the the boundary stress tensor.
Rotational invariance about the $x$ axis implies $dP_{\rm out}/d \Omega = 2 \pi \, d P_{\rm out}/ d \! \cos \theta$
where $\theta$ is the polar angle with $\theta = 0$ corresponding to the $+x$ direction the jet is moving.
In Appendix~\ref{sec:power} we compute the angular distribution
of power radiated by the jet exiting the slab.  The result reads 
\begin{equation}
\label{eq:stringpower2}
\frac{dP_{\rm out}}{d  \cos \theta} = \frac{1}{2} \int_{\sigma_*}^{\sigma_h(L)}  \! d\sigma 
\frac{-\pi_0^0(\sigma) }{\gamma(\sigma)^4 \left [1 {-} v(\sigma) \cos \theta \right ]^3 },
\end{equation}
where $v(\sigma)  = \partial_t x_{\rm geo} = \cos \tilde \sigma(\sigma)$ is 
the spatial velocity of the congruence of geodesics 
that make up the null string that exit the slab
and where $\gamma(\sigma) \equiv 1/\sqrt{1 - v(\sigma)^2}$ is the Lorentz boost factor.
Eq.~(\ref{eq:stringpower2}) shows how worldsheet energy $-\pi_0^0(\sigma)$ that exits the black hole slab is mapped 
onto the angular distribution of power on the boundary.
We note that for each $\sigma$, the integrand in Eq.~(\ref{eq:stringpower2}) is 
nothing more than 
a boosted spherical distribution of energy.  That is, boosting with velocity $-v(\sigma)$ in the 
$x-$direction, the integrand in Eq.~(\ref{eq:stringpower2}) becomes isotropic. 

Note that in the absence of any plasma we would have $\tilde\sigma=\sigma$ and 
the angular distribution of power would be given by (\ref{eq:stringpower2}) with
$v(\sigma)=\cos\sigma$, which is to say by $dP_{\rm in}/d \cos\theta$.

If {\it all} of the worldsheet energy $-\pi_0^0(\sigma)$ were localized at $\sigma=\sigma_*$,
Eq.~(\ref{eq:stringpower2}) would tell us that the ``jet'' in the boundary theory was 
a spherically symmetric cloud of energy with some energy $m$ in its rest frame --- i.e. in the
frame in which it is spherically symmetric --- that has subsequently been
boosted by a Lorentz boost factor $\gamma(\sigma_*)$.
The initial opening angle of the incident ``jet'' would be $m_{\rm in}/E_{\rm in}=\sin\sigma_*$
and the opening angle of the ``jet'' that emerges from the slab would be 
$m_{\rm out}/E_{\rm out}=\sin\tilde\sigma_*$.  We have seen that $E_{\rm out}<E_{\rm in}$
and $\tilde\sigma_* > \sigma_*$.  In both Fig.~\ref{fig:fallingstring} and Fig.~\ref{fig:morerelevant}
we find that $\tilde\sigma_*/\sigma_* > E_{\rm in}/E_{\rm out}$, meaning that $m_{\rm out}>m_{\rm in}$.
In fact we have found this to be the case in every example that we have investigated.

As long as $\sigma_*\ll 1$ the worldsheet energy density is in fact peaked near $\sigma=\sigma_*$, and 
the characterization that we have just given is a good approximation. This
characterization is not precise, however, because (\ref{eq:stringpower2}) describes
a ``jet'' composed by boosting spherically symmetric clouds of energy corresponding to 
the energy density at different $\sigma$ on the string worldsheet by different Lorentz boost factors.
The energy carried by the bits of string deeper in the bulk, at larger $\sigma$, is boosted less; it
describes the softer components of the ``jet''.

Let us now turn to the shape of the ``jet'' that exits the slab.
If we define its opening angle $\theta_{\rm out}$ as the angle
at which $dP_{\rm out}/d \cos\theta$ falls to one eighth of its peak (i.e. $\theta=0$)
value, 
inspection of (\ref{eq:stringpower2}) tells us that
\begin{equation}
\theta_{\rm out} \sim \tilde \sigma_*,
\end{equation}
as long as $\sigma_*\ll 1$ and as long as most of the worldsheet energy density
resides near $\sigma=\sigma_*$.
So, the angle at which the string endpoint falls into the bulk encodes how broad the ``jet'' 
is on the boundary.
Likewise, the opening angle of the incident jet is 
\begin{equation}
\theta_{\rm in} \sim \sigma_*.
\end{equation}
We know that $\tilde\sigma_*$ must be greater than $\sigma_*$:  in the dual gravitational, geometric,
description of jet quenching exemplified in Figs.~\ref{fig:fallingstring} and \ref{fig:morerelevant}
the slab of plasma is represented by the black hole horizon and its gravitational field, and this
gravitational field curves the trajectory of the string endpoint downward.  That is, $\tilde\sigma_*>\sigma_*$
because the force of gravity is attractive.  We now see that this basic feature of the bulk description of jet quenching
implies that $\theta_{\rm out}>\theta_{\rm in}$.  We can go a little farther upon
assuming that $x_{\rm stop}-L \gg u_h$ and $x_{\rm stop} - L = {\cal O}( x_{\rm stop})$.  
Under these assumptions, (\ref{eq:plasmageo}), (\ref{eq:xiout}) and (\ref{eq:xiright})  
yield
\begin{equation}
\tilde \sigma_* \sim \left ( \frac{u_h}{x_{\rm stop} - L} \right )^2.
\end{equation}
Hence $\theta_{\rm in} < \theta_{\rm out}  \ll 1$ as long as $x_{\rm stop} - L$ is much larger than
both $u_h$ and $|x_0|$.
That is, what comes out of the slab of plasma is a well collimated 
beam of energy until $L$ becomes parametrically close to $x_{\rm stop}$.

\begin{figure}[t]
\includegraphics[scale = 0.35]{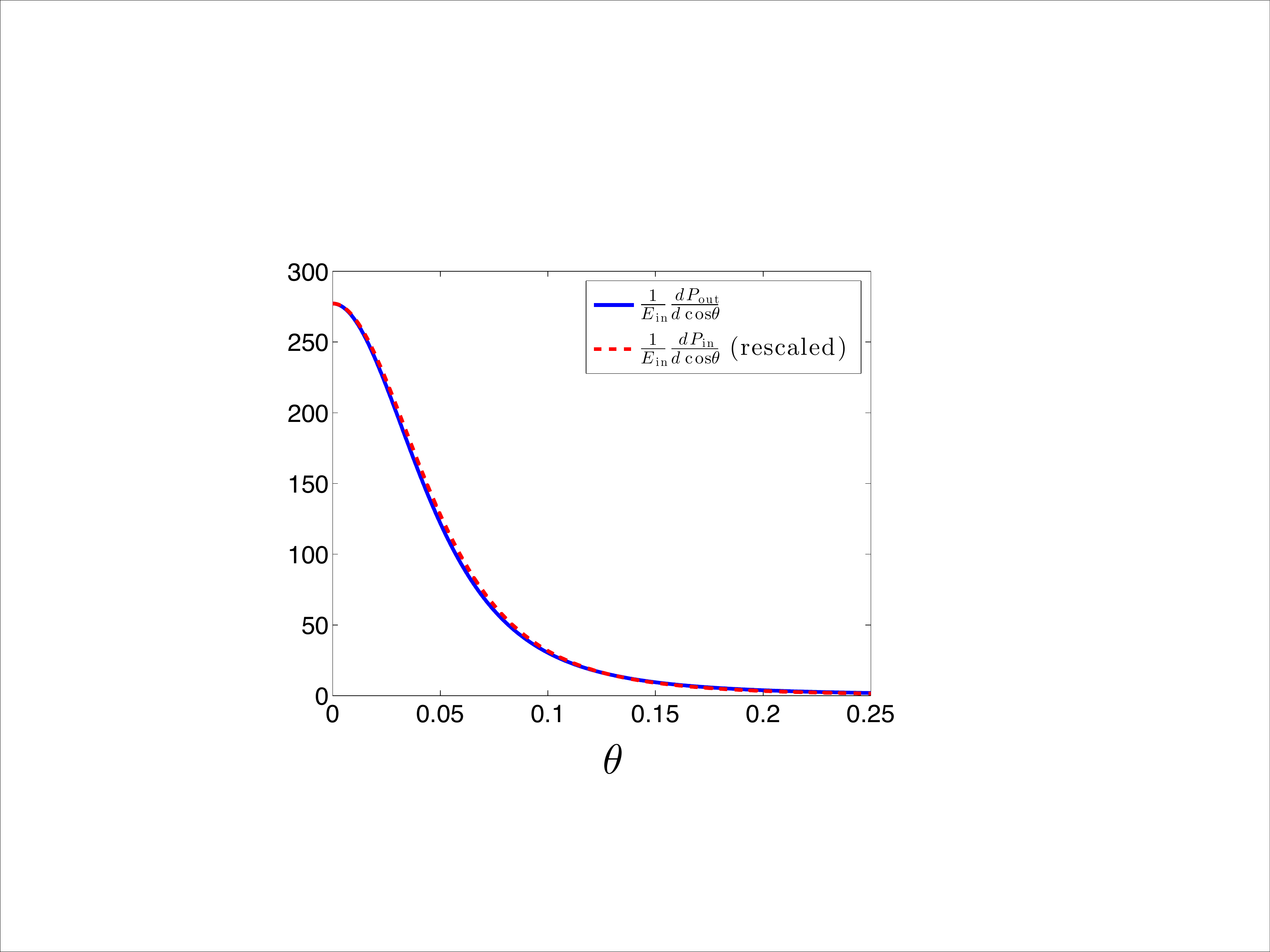}
\caption{ The angular distribution of power for the ``jet''
whose dual gravitational description is depicted in Fig.~\ref{fig:morerelevant} which
has traversed a slab of plasma with 
$L=8/(\pi T)$ and $x_{\rm stop}=10.73/(\pi T)$.
The blue solid curve shows 
$(1/E_{\rm in})(dP_{\rm out}/d\cos\theta)$.  We recall from Fig.~\ref{fig:morerelevant}
that $\tilde\sigma_*=0.0769$ and see here that the ``jet''
that emerges from the slab of plasma has an opening angle $\theta_{\rm out}$,
namely the angle at which the power has dropped to 1/8 of its $\theta=0$ value,
of this order.
We have also plotted the incident angular distribution of power 
$(1/E_{\rm in})(dP_{\rm in}/d\cos\theta)$, which is to say
the shape that the ``jet'' would have had in the absence
of any plasma, as the red dashed curve. In plotting the red dashed
curve we have stretched the $\theta$ axis by a factor of 3.2 and we have
compressed the vertical axis by a factor of 14.4.  
\label{fig:powers}
} 
\end{figure}

Fig.~\ref{fig:powers} shows the shape of the ``jet'' in the boundary quantum
field theory whose dual gravitational description is depicted in Fig.~\ref{fig:morerelevant}.
As is evident from the figure, the opening angle 
of $dP_{\rm out} / d \cos \theta$ is $\theta_{\rm out} \sim \tilde\sigma_* = 0.077$.  
Also shown 
in the figure is $dP_{\rm in}/ d \cos \theta$ with $\theta$ rescaled by a factor of 
3.2 and the amplitude rescaled by a factor of 1/14.4.  Aside from the rescalings,
we see that the shape of $dP_{\rm out}/ d \cos \theta$ is nearly identical to 
that of $dP_{\rm in}/ d \cos \theta$.  Therefore, just as the string that exits the black hole
slab looks identical to that which went in -- except with less energy and with an endpoint that falls with greater slope -- the angular distribution of power of the jet that exits the slab
is nearly identical in shape to that which went into the slab except its opening angle is larger and its energy has decreased.

From Fig.~\ref{fig:powers} we conclude
that the ``jet'' that emerges from the plasma is 3.2 times wider in angle than
the incident ``jet''.  We can compare this result to the
simpler estimate $\sin\tilde\sigma_*/\sin\sigma_*=3.08$ 
for the factor by which the opening angle should increase
that we obtained previously by assuming that the energy on the 
string worldsheet is localized near $\sigma=\sigma_*$.  The fact that
this simpler estimate is close to, but not equal to, the full boundary theory result obtained
in Fig.~\ref{fig:powers} tells us
that although the energy of the string worldsheet is peaked near $\sigma=\sigma_*$ 
it is not {\it all} localized there.

Let us now turn to energy loss in the slab.  
Integrating the angular distribution of power over all angles we find
\begin{equation}
\int d \! \cos \theta \frac{dP_{\rm out}}{d \! \cos \theta} = -  \int_{\sigma_*}^{\sigma_h(L)}  \! d\sigma \, \pi_0^0(\sigma)  = E_{\rm out}.
\label{eq:obvious}
\end{equation}
Therefore, the energy of the ``jet'' that exits the slab of plasma on the boundary coincides 
with the energy of the string which exits the black hole slab geometry in the bulk.
Likewise, the incident ``jet'' energy on the slab of plasma coincides with the incident string energy $E_{\rm in}$.
We see that by introducing a finite slab of plasma and asking about the energy of the ``jet''
that enters the slab and of the ``jet'' that exits the slab we find, by explicit computation, a completely straightforward
relationship between the ``jet'' energy in the boundary theory and the energy of the string
in the dual gravitational description, completely avoiding various ambiguities that
can arise in other contexts~\cite{Ficnar:2012np}.

\begin{figure}[t]
\includegraphics[scale = 0.35]{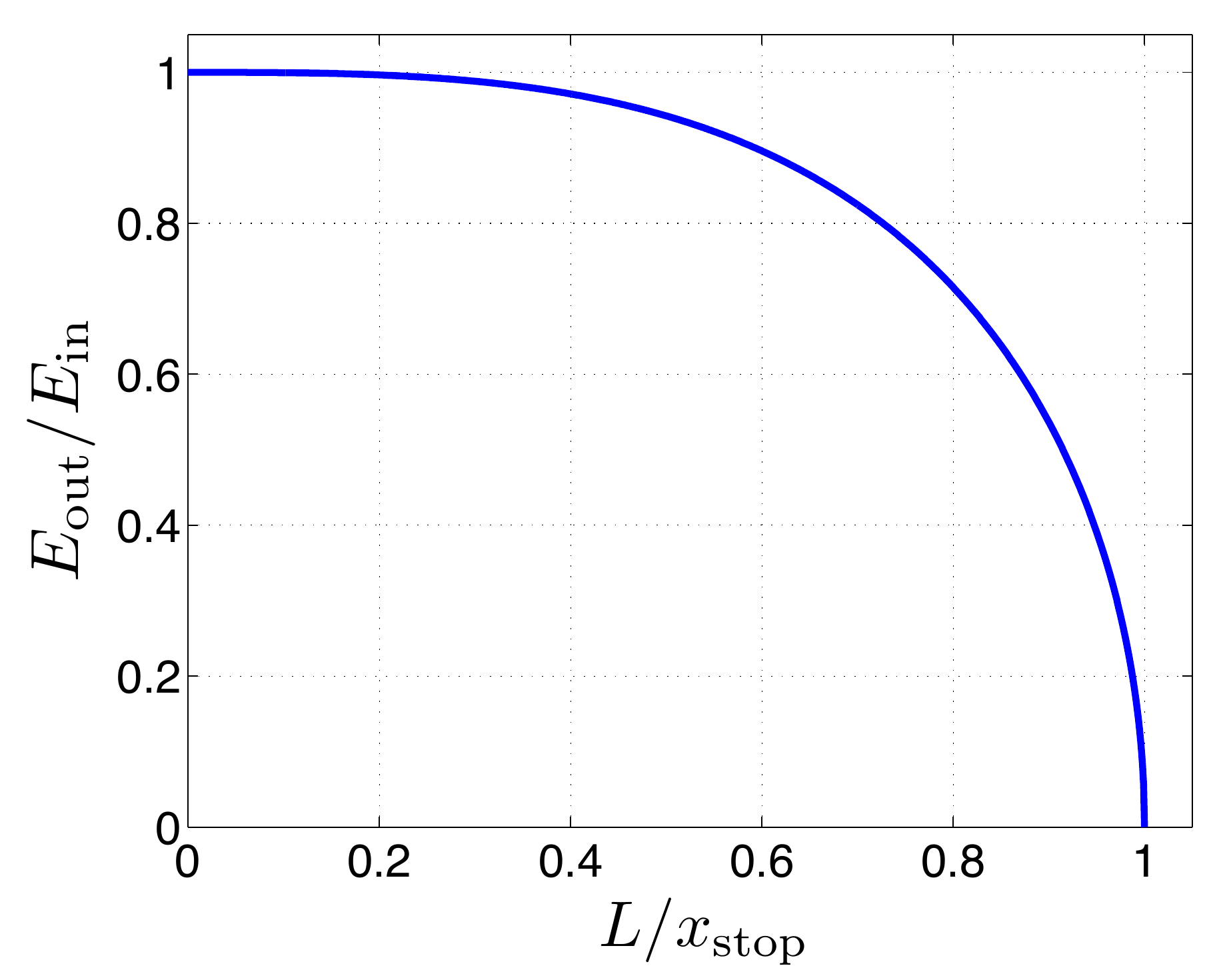}
\caption{ The ratio of  energies $E_{\rm out}/E_{\rm in}$ given in Eq.~(\ref{eq:Eratio}) 
as a function of $L/x_{\rm stop}$.
The energy loss rate $dE_{\rm out}/dL$ increases dramatically as $L \to x_{\rm stop}$.
This result for $E_{\rm out}/E_{\rm in}$ is accurate for any $x_0$ as long as $\sigma_*$
is small enough that $x_{\rm stop}\gg |x_0|+u_h$.  It provides a good approximation to the
energy loss of the ``jet'' depicted in Fig.~\ref{fig:morerelevant}.
\label{fig:Eratio}
} 
\end{figure}

We learn from (\ref{eq:obvious}) that the energy loss rate in Eq.~(\ref{eq:energylossrate}) and
the ratio $E_{\rm out}/E_{\rm in}$ in Eq.~(\ref{eq:Eratio}) that we obtained in the
previous section by computing  the energy of the string in Fig.~\ref{fig:morerelevant}
that enters, and exits, the slab of plasma does indeed give us the energy loss rate
and the ratio $E_{\rm out}/E_{\rm in}$ for the incident and outgoing ``jets'' in the boundary
quantum field theory.
We plot $E_{\rm out}/E_{\rm in}$ in Fig.~\ref{fig:Eratio}.  
We see from this figure that for $L= 0.5\,x_{\rm stop} $, $E_{\rm out} \approx 0.94\, E_{\rm in}$
and for $L= 0.9\, x_{\rm stop} $, $E_{\rm out} \approx 0.5\, E_{\rm in}$ 
and for $L= 0.98\, x_{\rm stop} $, $E_{\rm out} \approx 0.25\, E_{\rm in}$.  Therefore, as $L \to x_{\rm stop}$ the energy 
lost by the ``jet'' is disproportionately deposited near the end of its trajectory.
This is the signature of a Bragg peak energy loss rate for the ``jet'' in the plasma.  
In contrast to the conclusions reached in Ref.~\cite{Ficnar:2012np}, this demonstrates that the presence of the Bragg peak on the string 
worldsheet implies a Bragg peak on the boundary.  In would be interesting to do a full computation 
of the boundary stress tensor in the plasma in the vicinity of the Bragg peak.

\section{Outlook}
\label{sec:Outlook}

We have already stated our central conclusions in the introductory section of the paper. They
are demonstrated by Figs.~\ref{fig:fallingstring} and \ref{fig:morerelevant} which illustrate
the geometric interpretation of light quark energy loss in a strongly coupled plasma 
as due to null geodesics that carry energy along the string worldsheet falling into
the horizon
and which
show that even when the ``jet'' that emerges from the plasma has lost a substantial fraction
of its energy it looks precisely like the  ``jet'' that could have been
produced in vacuum with the same, reduced, energy $E_{\rm out}$
and the same, increased, opening angle $m_{\rm out}/E_{\rm out}$.
The latter conclusion is further reinforced in Fig.~\ref{fig:powers}.

We also note that the description  of the rate at which
a light quark loses energy as it propagates through strongly coupled plasma
that we have obtained in Eq.~(\ref{eq:energylossrate})
will be of use in many contexts.  It provides an expression
for $dE_{\rm out}/dL$ that can be used in the phenomenological
modeling of jet quenching in heavy ion collisions.  
It will also be interesting to analyze the consequences
for the analysis of jets in heavy ion collisions of
our result that $\theta_{\rm out}>\theta_{\rm in}$.
If, in the analysis of experimental data, the energy of a jet is defined as the energy inside
some specified opening angle, then if jets broaden in angle as they
traverse the quark-gluon plasma this could reduce their measured energy, over 
and above the ``true'' energy loss described by Eq.~(\ref{eq:energylossrate}).

The
expression (\ref{eq:energylossrate})  that we have derived
shows that
$|dE_{\rm out}/dL|\propto L^2$ for small $L$ and 
$|dE_{\rm out}/dL|\propto 1/\sqrt{x_{\rm stop}^2-L^2}$ for $L\sim x_{\rm stop}$,
with much of the initial energy of the jet lost near $L\sim x_{\rm stop}$
as in a Bragg peak and as illustrated in Fig.~\ref{fig:Eratio}.  We computed
the rate of energy loss given by (\ref{eq:energylossrate}) and illustrated in
Fig.~\ref{fig:Eratio} in the dual gravitational description of Fig.~\ref{fig:morerelevant}
by computing the energy of the string that emerges from the slab of plasma.
In Section~\ref{sec:BoundaryInterpretation} we confirmed by explicit calculation
that this is indeed the rate at which the ``jet'' in the boundary gauge theory
loses energy.

At a qualitative level, our observation in Figs.~\ref{fig:fallingstring}, \ref{fig:morerelevant}
and \ref{fig:powers} that the boosted beam of energy (the ``jet'') that
emerges from the plasma looks so similar in shape to the shape of the ``jets''
in vacuum in this theory resonates with the observations of jets in heavy ion
collisions at the LHC with which we began this paper.
We find that the propagation through the slab of plasma has two substantial
effects on the ``jets'' that we have investigated.  First, they lose energy,
as described by (\ref{eq:energylossrate}), as we have discussed.
Second, their opening angle increases.  
We find a simple geometric explanation
of the fact that the opening angle
$m_{\rm out}/E_{\rm out}$ after the ``jet'' traverses the plasma is always
greater than $m_{\rm in}/E_{\rm in}$:  in the dual gravitational description 
of jet quenching, this fact corresponds to the fact that gravity in 
the bulk ensures that the string endpoint
curves toward the black hole horizon.
In every example that we have investigated, we furthermore
find that $m_{\rm out}>m_{\rm in}$.

It remains the case that the ``jet'' that emerges
from the slab of plasma looks just like a ``jet'' in vacuum in the theory in which we 
are working.  
This is so because in this theory we can prepare a ``jet''
in vacuum with any value of $m/E$ that we like.
In QCD, on the other hand, the theory dictates the probability
distribution for $m_{\rm in}$ for jets with a given $E_{\rm in}$.
This jet mass probability distribution for both quark-initiated
and gluon-initiated jets has recently been
computed to next-to- and next-to-next-to-leading-log order in
Refs.~\cite{Dasgupta:2012hg,Chien:2012ur,Jouttenus:2013hs}.
It would be very interesting to construct an ensemble of ``jets'' in
the strongly coupled theory that we have employed with varying
values of $u_0$ and $\sigma_*$ such that the ensemble includes
jets with varying values of $E_{\rm in}$ and for each value of $E_{\rm in}$
includes varying values of $m_{\rm in}$ distributed as in QCD.  After
shooting this ensemble of ``jets'' through a slab of plasma one
could then look at the distribution of $E_{\rm out}$ and $m_{\rm out}$
for the ensemble of ``jets'' that emerge on the far side of the slab, for
example looking at the distribution of $m_{\rm out}$ for a specified $E_{\rm out}$.
Note that changing $m_{\rm in}$ at fixed $E_{\rm in}$ will change
both $E_{\rm out}$ and $m_{\rm out}$ meaning that in an investigation
like this it will be necessary to follow a two-parameter ensemble of ``jets''
through the slab.  We leave this investigation to future work.

It would of course also be interesting to replace the slab of plasma that
we have employed by an expanding cooling plasma that flows according
to the laws of hydrodynamics.  We leave this also to future work.

Another direction for the future is the tailoring of the ``jets'' in strongly
coupled ${\cal N}=4$ SYM theory so 
that they have the same shape as jets in QCD.   We have focused in this
paper on comparing the energy and shape of the  ``jets'' that emerge from the slab of plasma to
that of  the ``jets''
that are incident on it.  One could instead try to make a model for jets in QCD by replacing
(\ref{eq:endptenergy}) by an expression for $\pi_0^0(\sigma)$ tailored so that the angular
distribution of the energy in the ``jets'', see Fig.~\ref{fig:powers}, matches that of
jets in QCD.

Finally, it will be interesting to look 
for evidence in heavy ion collisions that
quenched jets have increased $m/E$ in addition to decreased $E$.  
Although it is difficult to measure the jet mass per se for jets in heavy
ion collisions, other jet shape observables have been measured~\cite{Chatrchyan:2013kwa}.
It would be interesting to analyze a sample of events each of which contains
a high energy photon with the same energy, with the photon back-to-back with jets of differing
energies in different events, to determine whether the jets that have lost more energy have
larger opening angles.  Present data sets~\cite{Chatrchyan:2012gt} 
do not include enough photon-jet 
events for such an analysis, but much higher statistics are anticipated in
coming years at the LHC.  It may also be possible to look for the effect on
a statistical basis in dijet events, looking for evidence that in asymmetric
dijets~\cite{Aad:2010bu,Chatrchyan:2011sx,Chatrchyan:2012nia} 
the lower energy jet in the pair has a larger angular extent.

\begin{acknowledgments}
We would like to thank Jorge Casalderrey-Solana, Doga Gulhan, 
Andreas Karch, Yen-Jie Lee, Hong Liu, Guilherme Milhano, Daniel Pablos, 
Iain Stewart and Jesse Thaler for helpful discussions.
KR is grateful to the CERN Theory Division for hospitality
at the time this research began.
This work was supported by the U.S. Department of Energy
under cooperative research agreement DE-FG0205ER41360.
The work of PC was also supported by the Fundamental
Laws Initiative of the Center for the Fundamental Laws
of Nature at Harvard University.

\end{acknowledgments}

\appendix

\section{The boundary angular distribution of radiated power}
\label{sec:power}

To compute the boundary angular distribution of power via (\ref{eq:powerdef}) we must
first compute the  linearized 
gravitational backreaction of the bulk geometry induced by the falling string.   The near-boundary behavior of 
the perturbations in the geometry then encode 
the expectation value of the boundary stress tensor $\langle T^{\mu \nu} \rangle$~\cite{deHaro:2000xn}.
Because $dP_{\rm out}/d\Omega$ only depends on the stress tensor asymptotically far from the slab, it is sufficient to 
study the perturbation in the AdS$_5$ geometry asymptotically far from the slab.
In other words, we can focus on the linearized backreaction of AdS$_5$ caused by 
the segment of string which exits the 
black hole slab.

The perturbation in the geometry due to the string is governed by linearized Einstein equations
sourced by the string stress tensor $\tau^{MN}$ given by 
\begin{equation}
\tau^{MN}(Y) =  \int d^2 \sigma \sqrt{-g} g^{ab} \partial_a X^M \partial_b X^N \frac{-T_0}{\sqrt{-G}} \delta^5(Y - X).
\end{equation}
With our choice of worldsheet coordinates, at leading order in the 
geometric optics expansion parameter $\epsilon$ the string stress tensor in the region $x > L$ reads 
\begin{align}
\nonumber
\hspace{-3mm} \tau^{MN}  = -\int_{\sigma_*}^{\sigma_h(L)} d \sigma  & \bigg [ u_{\rm geo}^2 \, \pi_0^0 \, \partial_t X^M_{\rm geo} \partial_t X^N_{\rm geo} \frac{1}{\sqrt{-G}} 
\\  \label{eq:stringstress}
 & \times \delta^{2}(\bm x_{\perp}) \delta(x - x_{\rm geo}) \delta(u - u_{\rm geo})  \bigg ].
\end{align}

The string stress tensor (\ref{eq:stringstress}) should be compared to that of a single point particle 
moving along a null geodesic $X_{\rm geo} = \{t,x_{\rm geo},0,0,u_{\rm geo}\}$, namely
\begin{align}
\nonumber
\tau^{MN}_{\rm particle} = & \,\varepsilon_o u_{\rm geo}^2 \gamma \partial_t X_{\rm geo}^M \partial_t X_{\rm geo}^N \\ 
&\times  \frac{1}{\sqrt{-G}} \delta^{2}(\bm x_{\perp}) \delta(x - x_{\rm geo}) \delta(u - u_{\rm geo}).
\label{eq:particlestress}
\end{align}
Here $\gamma \equiv 1/\sqrt{1 - v^2}$ with $v \equiv \dot x_{\rm geo}$
the velocity of the particle in the spatial direction.  $\varepsilon_o$ is a Lorentz scalar with respect  
to boosts in the boundary spatial directions.  In particular,
boosting to the frame in which $v = 0$, the energy of the null particle is 
simply $\varepsilon_o$.
Comparing (\ref{eq:stringstress}) and (\ref{eq:particlestress}) and noting that $\pi_0^0$ is time independent, 
we see that the string stress tensor is simply an integration over the congruence of null geodesics which make up the string, namely
\begin{equation}
\tau^{MN} = \int_{\sigma_*}^{\sigma_h(L)} d \sigma \, \tau^{MN}_{\rm particle}(\sigma),
\end{equation}
with a 
$\sigma$-dependent energy density 
\begin{equation}
\label{eq:sigmamass}
\varepsilon_o(\sigma) = - \frac{\pi_0^0(\sigma)}{\gamma(\sigma)},
\end{equation}
and a $\sigma$-dependent velocity $v(\sigma)$.
Linearity of the bulk to boundary problem then implies that the expectation value of the 
stress tensor induced by the string $\langle T^{\mu \nu} \rangle$ can be written as a sum
over that induced by null point particles following the (blue) null geodesics in a calculation
like that in Fig.~\ref{fig:fallingstring} or Fig.~\ref{fig:morerelevant}.  That is, 
\begin{equation}
\label{eq:bndstress}
\langle T^{\mu \nu}  \rangle = \int_{\sigma_*}^{\sigma_h(L)} d \sigma \, \langle T^{\mu \nu}_{\rm particle} \rangle.
\end{equation}
It therefore follows that 
\begin{equation}
\label{eq:powerfromparticles}
\frac{dP_{\rm out}}{d \Omega} = \int_{\sigma_*}^{\sigma_h(L)} d \sigma  \frac{dP_{\rm particle}}{d \Omega},
\end{equation}
with $dP_{\rm particle}/d \Omega$ defined by (\ref{eq:powerdef}) with the replacement $\langle T^{\mu \nu} \rangle \to \langle T^{\mu \nu}_{\rm particle} \rangle$

The boundary stress tensor induced by a single null particle falling in the 
AdS$_5$ geometry was computed in Ref.~\cite{Hatta:2010dz}.
Defining $x_{\rm bndy}^\mu$ as the event at which the geodesic starts from the boundary at $u = 0$,
the expectation value of the boundary stress tensor reads
\begin{align}
\label{eq:ptstress}
\langle T_{\rm particle}^{\mu \nu} \rangle = \frac{\varepsilon_o}{4 \pi r^2} \frac{1}{\gamma^3 ( 1 - \hat r \cdot \bm v)^3} \frac{\Delta x^\mu \Delta x^\nu}{r^2} \delta(t - t_{\rm bndy} - r),
\end{align}
where $\Delta x^\mu = x^\mu - x_{\rm bndy}^\mu$, $r = | \Delta \bm x|$ and $t = x_{\rm bndy}^0$.  
In the rest frame where $v = 0$ the induced stress on the boundary corresponds to a spherical shell 
of energy and momentum moving radially outwards from the event $x_{\rm bndy}^\mu$ at the speed of light.
We therefore have
\begin{equation}
\frac{dP_{\rm particle}}{d \Omega} =  \frac{\varepsilon_o}{4 \pi}  
 \frac{1}{\gamma^3(1 {-} \hat x \cdot \bm v)^3 }.
\end{equation}
Using (\ref{eq:powerfromparticles}) and (\ref{eq:sigmamass}) we therefore secure
\begin{equation}
\label{eq:stringtopower}
\frac{dP_{\rm out}}{d \Omega} =  \frac{1}{4 \pi} \int_{\sigma_*}^{\sigma_h(L)}  \! d\sigma 
 \frac{-\pi_0^0}{\gamma^4(1 {-} \hat x \cdot \bm v)^3 }.
\end{equation}
Upon multiplying by $2 \pi$ we  obtain the result (\ref{eq:stringpower2}) that
we have used throughout Section~\ref{sec:BoundaryInterpretation}.

\bibliography{refs}

\end{document}